# Two Novel Defences against Motion-Based Keystroke Inference Attacks

Yihang Song, Madhur Kukreti, Rahul Rawat, Urs Hengartner
Cheriton School of Computer Science
University of Waterloo
{y59song,mkukreti,r3rawat,urs.hengartner}@uwaterloo.ca

*Abstract*—Nowadays smartphones come embedded with multiple motion sensors, such as an accelerometer, a gyroscope and an orientation sensor. With these sensors, apps can gather more information and therefore provide end users with more functionality. However, these sensors also introduce the potential risk of leaking a user's private information because apps can access these sensors without requiring security permissions. By monitoring a device's motion, a malicious app may be able to infer sensitive information about the owner of the device. For example, related work has shown that sensitive information entered by a user on a device's touchscreen, such as numerical PINs or passwords, can be inferred from accelerometer and gyroscope data.

In this paper, we study these motion-based keystroke inference attacks to determine what information they need to succeed. Based on this study, we propose two novel approaches to defend against keystroke inference attacks: 1) Reducing sensor data accuracy; 2) Random keyboard layout generation. We present the design and the implementation of these two defences on the Android platform and show how they significantly reduce the accuracy of keystroke inference attacks. We also conduct multiple user studies to evaluate the usability and feasibility of these two defences. Finally, we determine the impact of the defences on apps that have legitimate reasons to access motion sensors and show that the impact is negligible.

## I. INTRODUCTION

Almost all of the smartphones shipped today include a number of sensors, such as an accelerometer, a gyroscope, an orientation sensor, a barometer, or rotational vector sensors. All these sensors provide raw data with high precision and accuracy. The data collected from these sensors can be put to a number of uses. Monitoring three-dimensional device movement or positioning, monitoring changes in the environment near a device, or motion-based commands are a few examples.

The data collected by these sensors is not treated as sensitive data by Android, iOS, and Blackberry. Third-party apps across all three platforms are allowed to access the sensor readings without any security permission requirements.

Since smartphones are shipped with numerous embedded sensors, concerns have been raised about the ways in which unrestricted access to sensor readings poses a potential threat to a user's private information. The keyboard is the most widely used input device today. Lots of sensitive information, such as passwords or credit card numbers, are typed using a keyboard. Most of the smartphones today do not have a physical keyboard. The user is instead provided with an on-screen software keyboard. Traditional keyloggers face an obstacle because all operating systems allow only those apps to read keystrokes that are active and have the focus on the screen. As a result, new ways have been devised to overcome this obstacle and infer the keystrokes that a user makes on the device.

Several researchers [9][10][12][15][16][17] have shown how motion-based side-channel attacks can be used to infer the keys typed by a user. For example, Xu et al. [17] present TapLogger, which uses data collected from the accelerometer and the orientation sensor to infer a user's input. By observing the relation between tap events and the motion change of a device, TapLogger is able to infer a user's input, namely a numerical PIN or a password, with high probability.

In this work, we present two defences against motion-based keystroke inference attacks. Our work is based on two observations. First, the attacks make use of accelerometer data to segment between key strokes before they infer the type of key stroke. We present a method to alter the accelerometer readings just before they are passed to third-party apps to ensure that a key stroke event is not detected by the attacker. Second, the attacks work under the assumption that a particular key is always displayed in a certain position on the screen. Our second defence approach is to use a randomized keyboard layout that is different from a traditional keyboard layout in the sense that each key is in a different position every time the keyboard is displayed on-screen.

For the first defence mechanism, the modified accelerometer readings, we examine the impact of these modifications on key stroke detection. We also perform a user study to document whether a user is able to perceive any difference in apps that use the accelerometer after these changes are made. We present our second defence mechanism, the randomized keyboard layout app, to users to determine the balance of usability and security that a user desires to achieve. Finally, we evaluate whether an app that needs the accelerometer for legitimate purposes is affected by these changes.

The rest of this paper is organized as follows: Section 2 provides some fundamental information about Android and the related sensors. Section 3 follows with a brief presentation of some existing motion-based keystroke inference attacks. Section 4 demonstrates our defences in details and Section 5 gives our implementation details. In Section 6, we evaluate our defences and present the results and analysis. Some limitations are mentioned in Section 7. We come up with some possible future work in Section 8 and conclude in Section 9.

```xml
<Keyboard
        android:keyWidth="%10p"
        android:keyHeight="50px"
        android:horizontalGap="2px"
        android:verticalGap="2px" >
    <Row android:keyWidth="32px" >
        <Key android:keyLabel="A" />
        ...
    </Row>
    ...
</Keyboard>
```

Fig. 1: A sample keyboard layout file.

## II. TECHNICAL BACKGROUND

### A. Keyboard Input on Smartphones

As stated earlier, most of the smartphones today do not have a physical keyboard. This means that a software keyboard displayed on the touchscreen is used instead as a user interface on a smartphone. Every tap event on the screen is understood by the smartphone OS as the coordinates of the position on the screen where the user tapped. The user input is then inferred using the known coordinates and knowledge of the app view currently being displayed on the screen. User input using the keyboard can be viewed as a series of tap events where each tap event with coordinates within the boundary of a displayed button represents a tap on that particular button. The layout used for text input is the standard QWERTY layout across all platforms. The other used layouts, such as the lock screen keypad, are public knowledge and uniform.

Android provides protection against traditional keyloggers by restricting access to the coordinate information of the tap events. Only the view that is focused and being displayed on the screen is capable of receiving the tap events and the related coordinate information [1]. A keylogger that runs as a third-party app in the background is not allowed access to this data and thus cannot infer where the tap event took place based on this data. However, any app can receive raw data from motion sensors (e.g. accelerometer, gyroscope) because this data is not treated as sensitive by the OS. The objective of motion-based keystroke inference attacks is to infer user input without any knowledge of coordinate data available to the app view and only on the basis of raw motion sensor data resulting from motion changes of the smartphone due to tapping on the touchscreen.

### B. Keyboard in Android

A keyboard in Android consists of rows of keys. The public class `Keyboard`, which extends the class `Object`, loads an XML description of a keyboard and stores the attributes of the keys [2]. The XML description is stored in a layout file that looks like the snippet of code in Figure 1. The layout file contains the XML attributes listed in Table I.

Each attribute in Table I defines how the on-screen keyboard is to be displayed. The class `Keyboard.Key` describes the position and characteristics of a single key on the keyboard.

TABLE I: XML attributes of the keyboard layout file.

| Attribute Name | Description |
|---|---|
| android:horizontalGap | default horizontal gap between keys |
| android:keyHeight | default height of a key, in pixels or percentage of display height |
| android:keyWidth | default width of a key, in pixels or percentage of display width |
| android:verticalGap | default vertical gap between rows of keys |

The static class `Keyboard.Row` is a container for keys on the keyboard.

### C. Sensors in Android

The Android platform provides support for three categories of sensors [3]:

- Motion Sensors: This category includes accelerometers, gravity sensors, gyroscopes, and rotational vector sensors. They are used for measuring acceleration and rotational forces along the x, y and z axis.

- Environmental Sensors: Sensors such as barometers, photometers, thermometers measure environmental factors such as the temperature, pressure, or humidity.

- Position Sensors: Orientation sensors and magnetometers measure the physical position of the mobile device.

Sensors are either hardware-based or software-based. Hardware-based sensors are physical components embedded into the device and derive their data by measuring environmental parameters. Software-based sensors, on the other hand, derive their data from the on-board hardware-based sensors. For example, the (deprecated) software-based orientation sensor derives its measurements from accelerometer data.

The Android sensor framework comprises of several classes and interfaces that enable a developer to perform sensor-related tasks. Among these, acquiring raw sensor data and registering sensor event listeners that monitor changes in sensor readings are the ones that are significant in the context of using sensor data for inferring user input. Namely, the sensor framework consists of the following classes and interfaces [3]:

- `SensorManager` This class is used to create an instance of a sensor service. It provides the methods for registering sensor event listeners and acquiring orientation information among others.

- `Sensor`: This class is used to create an instance of a particular sensor and includes methods to determine that sensor's capabilities.

- `SensorEvent`: This class creates a sensor event object that provides information about a sensor event. Sensor event objects consist of the following pieces: the raw sensor data, the type of sensor that generated the event, the accuracy of the data, and the timestamp for the event.

- `SensorEventListener`: This interface provides two methods `onAccuracyChanged()` and `onSensorChanged()` that receive sensor events

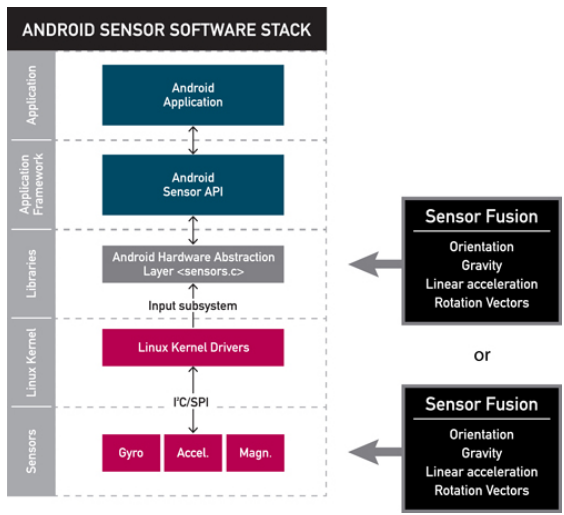

Fig. 2: Android sensors framework [8].

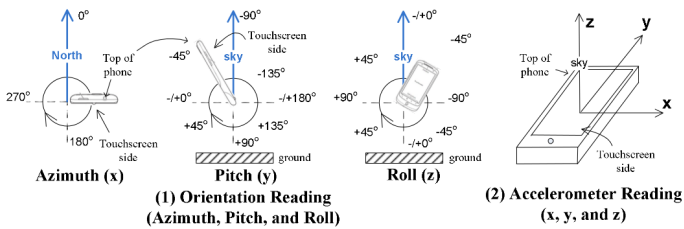

Fig. 3: Accelerometer and orientation sensor [17].

when the sensor accuracy changes and the sensor values change, respectively.

Figure 2 shows an overview of the Android sensors framework. The control flows from an app that registers an `EventListener` with the `SensorManager` towards the drivers that collect raw data from the sensors. The data flows from these device drivers to the app. There is no restriction on who can register a listener and receive this raw data. This information is not treated as sensitive and thus exposes a user to malicious apps that try to infer keystrokes.

Figure 3 illustrates readings from the accelerometer and the orientation sensor on the Android platform.

### III. MOTION-BASED KEYSTROKE INFERENCE ATTACKS

Several researchers [9][10][12][15][16][17] have shown how motion-based side-channel attacks can be used to infer the keys typed by a user. The general approach of these existing motion-based keystroke inference attacks is similar. Basically, an attack consists of two steps. The first step is *tap detection*, that is, finding the start and end of a user tap in the sensor data. The second step is *tap inference*, that is, inferring the pressed key based on the sensor readings in that period. Most of the existing attacks focus on the second step. They execute this step by extracting features from the sensor readings and using machine learning techniques to infer the key. More details about each existing attack are given in Section VIII. In the remainder of this section, we focus on TapLogger.

Xu et al. [17] give the design and implementation of TapLogger, a Trojan app that infers a user's typed numerical PINs and passwords by using data from the accelerometer and the orientation sensor. Because this paper provides details about both tap detection and tap inference and because we also managed to get part of the source code from the authors, we use TapLogger to illustrate the design of an app that executes a motion-based keystroke inference attack.

TapLogger works in two modes: Training mode and logging mode.

In the training mode, readings of the accelerometer and the orientation sensor are used to generate a user interaction pattern. In this mode, the user interacts with the TapLogger app and TapLogger legitimately receives the tap events and their coordinates. These coordinates along with the raw sensor data are the input that TapLogger uses to generate the user interaction pattern. In this mode, for each tap event TapLogger records the coordinates, the timestamp of the beginning and end of a tap event and obviously the raw sensor data from the accelerometer and the orientation sensor.

In the logging mode, a user is interacting with an app that requires the user to enter sensitive information, such as passwords or PINs, using the on-screen keyboard. TapLogger registers a listener with the `SensorManager` and receives the sensor readings. These readings along with the interaction pattern developed in the training mode are used to infer the user input without actually receiving coordinates for the tap events.

Unfortunately, the source code we got from the authors contains only the code for tap detection and for logging sensor data. No code for tap inference, that is, training, feature extraction, or classification, is included. Therefore, we implemented these components ourselves based on the description in the paper. However, we failed to get accuracy results in the range of the results reported in the paper based on this implementation. Therefore, we had to find other features. These features include (see Figure 4):

- F1: Maximal value during the tap
- F2: Minimal value during the tap
- F3: The index of the maximal value
- F4: The index of the minimal value
- F5: The difference between the last and the first value

All features apply to both pitch (y-axis) and roll (z-axis) data. The reason we do not use the x-axis value is because its relation to the type of tapped key is limited.

To evaluate our implementation, we use probability prediction provided by libSVM [6] and determine the top 1 inference and the top 4 inferences, which are the most probable inferred key stroke and the four most probable inferred key strokes, respectively. When evaluating the new features on a number pad (three rows and four columns), we manage to achieve 50% accuracy for the top 1 inference (i.e., in 50% of the cases the inferred key corresponds to the actually tapped key) and more than 80% accuracy for the top 4 inferences (i.e., the top 4 inferred keys contain the actually tapped key in 80% of the cases). Note that the focus of this evaluation is on tap inference, not tap detection, so we assume that tap detection is perfect and use the actual touch events to detect the start and end time of a tap.

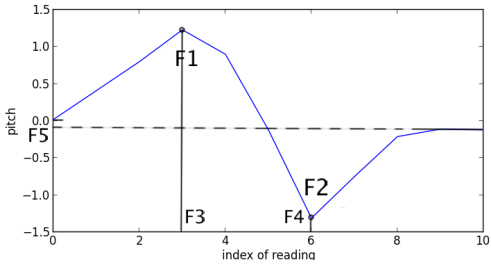

Fig. 4: Features of orientation data for tap inference.

Our tap inference code is publicly available.[1]

## IV. DESIGN OF DEFENCES

As described in the previous section, a motion-based keystroke inference attack generally consists of two phases: tap detection and tap inference. Intuitively, if we can block either of these two phases, it is possible to defend against such attacks. We now present two defences, the first one targeting tap detection, the second one targeting tap inference.

### A. Reducing the Accuracy of Sensor Data

*1) Approach:* The first defence is based on reducing the accuracy of sensor data given to apps that request this data. The reduction must be of a granularity such that apps that legitimately access this data preserve their usefulness but malicious apps that try to execute a motion-based keystroke inference attack will fail to execute this attack. This defence could be used against the tap detection phase or the tap inference phase of the attack. Of course, successful tap detection is a requirement for tap inference so if tap detection fails, the entire motion-based keystroke inference attack will fail.

As it turns out, applying the defence to tap inference would be harder than applying it to tap detection since different researchers have managed to infer taps from different sensors. Therefore, a comprehensive defence would have to reduce the accuracy of multiple sensors. Again, for each sensor we would also have to ensure that apps that have legitimate access to the sensor remain useful. On the other hand, tap detection generally is based only on data from the accelerometer. Namely, when closely examining the source code for the only two pieces of related work for which we managed to get access to source code, TouchLogger [11] and TapLogger [17], we noticed that both of them use the accelerometer for tap detection. In addition, ACCessory [16] and Aviv et al. [10] exclusively use the accelerometer for their motion-based keystroke inference attacks. Therefore, a defence based on reducing the accuracy of sensor data during tap detection needs to take only the accelerometer into account.

As mentioned before, TapLogger is the most detailed when it comes to describing tap detection. Namely, TapLogger detects a tap based on change of acceleration, as shown in Figure 5. The extracted features, $P_1, ..., P_5$, characterize the change of the square sum of the acceleration in three directions, $SqSum$. The change of $SqSum$ is always in a small range during tapping. Therefore, if we can hide these small changes from an app, a malicious app can no longer detect a tap. On the other hand, large changes will not be hidden from an app, so apps that have legitimate access to acceleration data should continue to work.

[1]https://bitbucket.org/Near/taploggertrainer

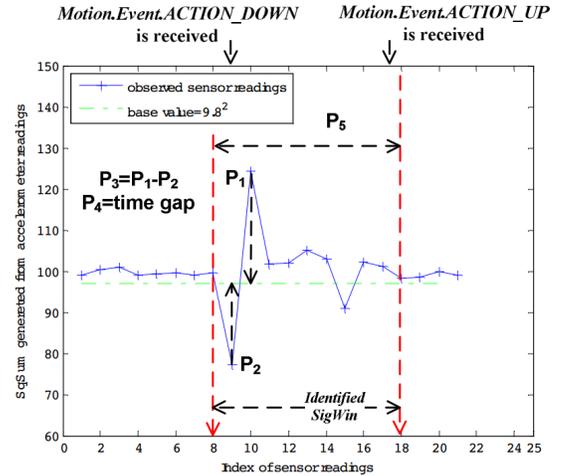

Fig. 5: Features of acceleration data for tap detection [17].

The scenario described above assumes that the device is relatively stable while the user is tapping on it. However, when the device is not stable, for example, because the user is walking while tapping, the change of the square sum will likely not be in this small range so our defence will not hide the changes from an app. However, the amount of noise caused by the device not being stable will drown out the small changes from tapping so tap detection will fail.

Note that applying the idea of hiding small changes from apps to sensors other than the accelerometer may be difficult. For example, small changes reported from the orientation sensor can have a big impact on apps that legitimately use this sensor because the base is also low.

*2) Challenges:* This approach needs modifications at the kernel level, thus it becomes harder to be widely deployed.

Reducing the accuracy may introduce usability issues. Some apps that use the accelerometer for legitimate reasons may require very high accuracy in sensor readings. As a result, these apps may not work well with our defence modifications. It is hard to set the degree of reduction such that we can defend against motion-based keystroke inference attacks without affecting other apps. We will evaluate this issue in our user studies.

### B. Random Keyboard Layout Generation

*1) Approach:* During the tap inference phase of a motion-based keystroke inference attack, the attacker tries to infer the position of a tap on the screen. The actual pressed key can then easily be learned since the layout of the keyboard is typically public and therefore known to the attacker. The idea behind our second defence is to hide this layout information from the attacker by randomizing the keyboard layout.

*2) Challenges:* Fundamentally, a random keyboard layout is just a form of substitution cipher, which is vulnerable to frequency analysis attacks. Some confidential content that users input in their mobile devices, such as passwords, may be short and irregular, which makes them less vulnerable to frequency analysis attacks. However, frequency analysis attacks are definitely a concern for other types of sensitive input, such as a confidential email or text message. To defend against frequency analysis attacks, we need to shuffle the keyboard layout again and again to avoid using the same substitution "key" for a long time.

Besides, with the random keyboard layout, users cannot type as fast as usual because the keys are not where they are on a regular keyboard. Therefore, there is a trade-off between usability and confidentiality. To evaluate the degree of influence on the typing speed, we also conduct a user study.

### C. Other Possible Defences

There are quite a few possible countermeasures mentioned in other papers, like lowering the sampling rate of sensors [10][15][16], adding permissions for using the sensor data [15][16], denying access to sensors while the user is inputting sensitive data [10][15], vetting apps that access sensor data [10], or using a leather or rubber case that absorbs the device motion [15].

We considered using an approach based on lowering the sampling rate of a sensor instead of reducing its accuracy, but decided against it due to observations made in earlier work [15], indicating that lowering the sampling rate would greatly affect the usefulness of apps with legitimate reasons to access a sensor.

Some of the proposed methods require a good understanding of security from users, which is not necessarily the case. Namely, adding permissions for using the sensor data would require users to read and understand an app's permission dialog before installing it, which may not happen in practice [13][14].

Denying access to sensors while the user is inputting sensitive data has the potential to break some apps. For example, a pedometer app would break if a user entered text while walking.

Vetting apps that access sensor data may work for some platforms that are already based on vetting, like iOS, but does not work for platforms without a formal vetting process, like Android. Of course, a vetting-based approach would fail for rooted devices, where users decide themselves which apps to install.

Using a special case protects only the subset of the users that bother acquiring such a case, whereas we are looking for a solution from which all users can benefit without having to go through additional expenses.

Our proposed two defences are relatively easy to deploy. Reducing the accuracy does require a kernel change (as would lowering the sampling rate), but no changes to apps are required (as would introducing new permissions). Similarly, no changes to apps are required for the randomized keyboard layout. Currently, a user who wants to benefit from this defence only needs to install the corresponding input method and switch to it when inputting sensitive information, which is very easy in Android. Therefore, this defence has the advantage that, as opposed to most other proposed defences, it can be used by security-conscious users right away. Of course, the defence may become integrated with a phone platform over time.

Our defences require no or very little security understanding from users. The first defence is always enabled, regardless of a user's actions. Similarly, a randomized keyboard layout could be used for any user input. However, for usability reasons, we suggest that a mobile phone platform uses such a layout only when passwords or PINs are entered. In addition, the platform could give the user the option to switch to a randomized layout while entering other types of sensitive text.

## V. IMPLEMENTATION

We implemented our two defences on the Android platform because Android is a very popular platform for mobile devices and is also easy to use by app developers. Besides, reducing the accuracy of sensor data involves some modifications to the kernel. Android, which is open source, is therefore a good choice for implementing a prototype.

### A. Reducing the Accuracy of Sensor Data

To lower the accuracy of sensor data, we have to make some modifications to the Android operating system. Sensor data is sent to a device, and the JNI component will read the data from the device file, just as in the basic Linux kernel. In the Java part, the `SensorManager` class will call the native function provided by JNI to get the sensor data. When an app registers a listener, the `SensorManager` will save it to a list. Every time a sensor event occurs, it will pack the data into a sensor event object and apply the event handler defined by a developer in the listener to the object.

We added some code just before the handler function is called. For an accelerometer event, the sensor event object will contain three values, which represent the acceleration in three directions. As mentioned in Section IV-A, TapLogger uses the change of the square sum of these three values to detect a tap. We observe that for a normal tap, the square sum is always in the range of $(80, 130)m^2/s^4$, so for every event in this range, we will maintain the direction of the acceleration but set its square sum to a constant value, namely, $9.8*9.8m^2/s^4$. Thus TapLogger will fail to detect almost all taps.

For simplicity, we lower the accuracy of sensor data for all apps requesting this data. In a more sophisticated solution, we could lower the accuracy only for background apps but leave it unchanged for the foreground app.

### B. Random Keyboard Layout Generation

For the generation of a random keyboard layout, we develop a new input method app, which is based on the official example app given by Google. A normal input method app will first load the layout specified in an XML file and then generate an inner representation of this layout, which is actually an array

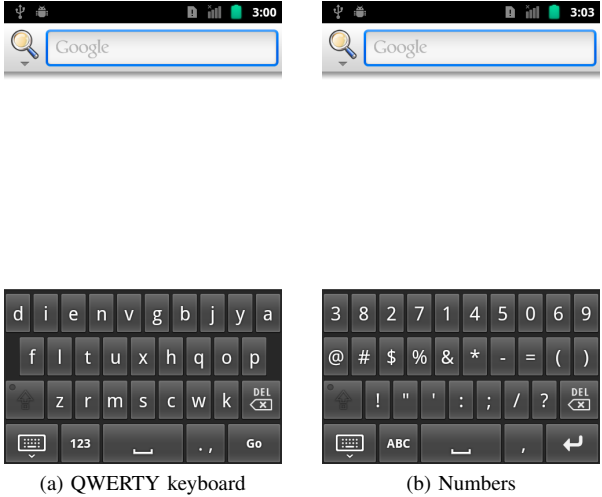

(a) QWERTY keyboard  (b) Numbers

Fig. 6: Implementation of random keyboard layout.

of keys. Each key object will contain the position, the size, and the corresponding keycode.

To achieve our goal, we have to solve a few problems. First, we need to randomly permute the keys. Because the keys near the boundary of the keyboard are slightly different from other keys in the XML layout file (they have different attributes), instead of generating a random XML file and loading it, we shuffle the array of keys only after loading the XML file. Additionally, in this way we need to load the XML file only once and introduce negligible overhead. The random permutation of an array is quite easy. For every element, we pick an element after it randomly and swap the two elements.

Second, because we do not want to use the same substitution continuously, which would be vulnerable to frequency analysis attacks, we generate a different random layout every time the keyboard pops up. In particular, we analyzed the workflow of the input method service and then added the permutation operation to the function that shows the keyboard.

Figures 6a and 6b show an example of what the user will see when using a random keyboard layout.

## VI. EVALUATIONS

### A. Impact of Accuracy Reduction on Tap Detection

In our first experiment, we determine the impact of reducing the accuracy of the accelerometer data on tap detection. We have a user enter 300 keystrokes on an unmodified Nexus One device and on a Nexus One device that reduces accuracy. For tap detection, we use the source code received from the authors of TapLogger.

For the unmodified device, 120 keystrokes are detected and 60 false positives are found. For the modified device, 50 keystrokes are detected and 8 false positives are found. Therefore, our defence increases the detection rate more than 50%.

We find that the taps that are still detected in spite of our defence are those where the user applies more force, therefore their square sum of the acceleration is not in the range that we suppress. It is possible to enlarge this range at the cost of affecting apps that legitimately use accelerometer data.

For our current choice of parameters, only about one out of six taps will be detected. This significantly increases the attacker's challenge when it comes to trying to infer a user's password or PIN. Similarly, we expect that it becomes much harder to infer the content of an email or SMS message if only one sixth of the content is available and where the still available content may have been inferred incorrectly due to the accuracy limits of tap inference. Studying the difficulty of inferring this kind of content is future work.

### B. Impact of Accuracy Reduction on Other Apps

In the next two experiments, we investigate the impact of reducing the accuracy of the accelerometer data on two apps that legitimately access the accelerometer.

*1) Experiment 1:* We conducted a study on ten Android users from our university. The participants were required to have basic comfort with touchscreens and motion gaming. The participants were required to play three stages of "Crazy Labyrinth 3D" [5] on an unmodified Nexus One device and on a modified Nexus One device. The participants had no knowledge about the different operating systems on the devices. A device was chosen randomly in the beginning and the devices were alternated after successful completion of every stage. Completion time for each stage was recorded for both devices. The purpose of this study is to gauge the effect of reducing the accuracy on apps that require sensor data. Inaccurate sensor readings may lead to malfunctioning in such apps and hence we wanted to analyze the behaviour of such apps on our modified Android system. Crazy Labyrinth 3D was chosen due to its complete dependence on the motion of the device so any fluctuations in sensor readings would directly affect the gameplay. Table II shows the results.

Although we modify the absolute value of the acceleration, we maintain the direction. We believe that the game mainly uses the direction to set the move of the ball. The interesting observation is that users took more time to complete the three stages on the unmodified device than on the modified device. This difference can be attributed to the fact that lower sensor accuracy must have also reduced the noise in the readings hence giving users better control over the movement of the ball in the game.

The participants then answered a small questionnaire regarding their experience on the two devices. Namely, the participants were required to give their device preference based on their gameplay experience.

From Table III, we find that most of the users were either not able to spot any difference between the devices or preferred the modified device over the unmodified one.

*2) Experiment 2:* We also test how our kernel modification affects a step counting app. We choose Pedometer [7], a well known step counting app. We run this app on both the unmodified and modified devices and walk 100 steps. Figures 7a and 7b show the result. From these two pictures, we can see that there are only two steps difference between these two devices and we can confidently claim that this modification will not affect step counting apps.

TABLE II: Time taken for playing games.

| Participants | Stage 1 (sec) | | Stage 2 (sec) | | Stage 3 (sec) | |
|---|---|---|---|---|---|---|
| | Unmodified | Modified | Unmodified | Modified | Unmodified | Modified |
| 1 | 7.143 | 4.170 | 7.402 | 5.179 | 8.529 | 7.178 |
| 2 | 6.374 | 7.944 | 5.448 | 4.630 | 11.417 | 10.072 |
| 3 | 5.767 | 9.441 | 5.567 | 5.215 | 7.981 | 6.056 |
| 4 | 3.345 | 5.347 | 5.412 | 4.551 | 7.064 | 6.191 |
| 5 | 9.37 | 5.344 | 6.082 | 5.52 | 8.934 | 5.221 |
| 6 | 7.834 | 6.185 | 6.567 | 6.765 | 10.256 | 8.126 |
| 7 | 7.453 | 7.165 | 7.190 | 5.224 | 11.327 | 7.479 |
| 8 | 5.236 | 6.437 | 6.982 | 5.115 | 8.537 | 7.462 |
| 9 | 7.134 | 5.784 | 6.952 | 5.564 | 8.247 | 6.225 |
| 10 | 6.132 | 7.155 | 6.547 | 7.126 | 8.147 | 9.478 |
| Mean | 6.578 | 6.497 | 6.414 | 5.488 | 9.04 | 7.348 |

TABLE III: Questionnaire.

| Participant # | Which device is better? |
|---|---|
| 1 | No difference |
| 2 | Modified |
| 3 | Modified |
| 4 | No Difference |
| 5 | No Difference |
| 6 | No Difference |
| 7 | Modified |
| 8 | Unmodified |
| 9 | No Difference |
| 10 | No Difference |

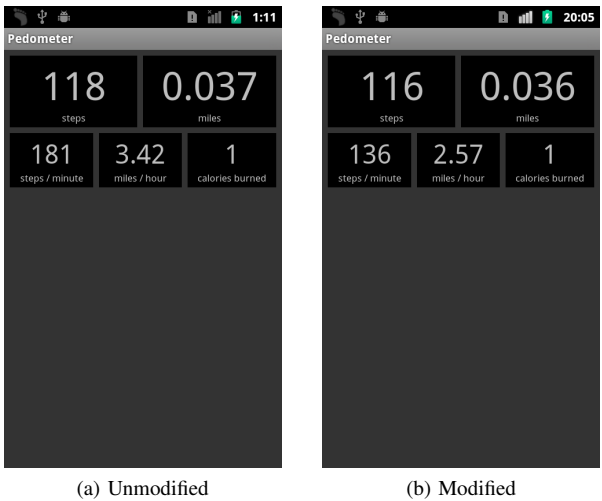

(a) Unmodified   (b) Modified

Fig. 7: Step counting result.

*C. Impact of Random Keyboard Layout on Input Time*

In this experiment, we determine how a random keyboard layout affects the time to enter input. We used the same ten participants as in the earlier experiment to answer this question. The participants were required to type the following two messages:

- Message 1: This is a message. I would like to type the message.
- Message 2: thisismypassword123

The subjects first typed these two messages on the normal keyboard layout and then on a randomized keyboard layout. Their time to enter the messages was recorded. Capitalization was not required, however, users were required to use punctuation. The purpose of this experiment is to determine the feasibility and usability of a randomized keyboard layout for a regular user. While Message 1 is similar to our everyday SMS/Email text with spaces and punctuation, Message 2 represents passwords and other sensitive information, where white spaces are generally not necessary and a combination of characters and numbers is required.

As we can see from Table IV, the time for typing the message with a randomized keyboard layout is almost twice the time than the time taken with the regular one. We also observe that the difference between the mean typing times of Message 2 for the modified and the unmodified device is 18 seconds whereas in case of Message 2 it is 39 seconds. The time difference is proportional to the length of the typed messages. Based on these results, a randomized keyboard layout is feasible for short messages where security is really important, like a password, but we do not suggest the use of a randomized keyboard layout for long messages. Instead, a solution where the user is given the option to temporarily use a randomized keyboard layout while entering sensitive information, like a credit card number, as part of a long message is more attractive.

At the end of the experiment, we informed the participants about the purpose of the experiment and about motion-based keystroke inference attacks. We also asked the participants whether they were willing to use a randomized keyboard layout as a defence against these attacks. All participants indicated that they were willing to use a randomized keyboard layout while inputting sensitive information.

## VII. LIMITATIONS

Even though reducing the accuracy of sensor data and randomizing the keyboard provide effective protection against keystroke inference attacks, there are some limitations to our defences.

TABLE IV: Time taken for typing messages.

| Participants | Message 1 (mm:ss) | | Message 2 (mm:ss) | |
|---|---|---|---|---|
| | Regular | Randomized | Regular | Randomized |
| 1 | 00:28 | 01:01 | 00:10 | 00:25 |
| 2 | 00:27 | 00:47 | 00:09 | 00:21 |
| 3 | 00:24 | 01:13 | 00:11 | 00:25 |
| 4 | 00:18 | 01:28 | 00:09 | 00:38 |
| 5 | 00:23 | 01:06 | 00:07 | 00:32 |
| 6 | 00:20 | 00:49 | 00:10 | 00:25 |
| 7 | 00:22 | 00:59 | 00:11 | 00:28 |
| 8 | 00:20 | 00:53 | 00:10 | 00:27 |
| 9 | 00:21 | 00:57 | 00:11 | 00:24 |
| 10 | 00:19 | 01:07 | 00:10 | 00:30 |
| Mean | 00:23 | 01:02 | 00:10 | 00:28 |

First, the defences were only checked for one game (i.e., Crazy Labyrinth 3D) and one app (i.e., Pedometer). There may be apps in the Android market that require more precise and accurate sensor readings than these two apps.

Second, the number of users in our experiments was rather small. Results based on a small user set may not reveal the user experience and perception in their entirety.

Third, the accuracy reduction of the sensor data may depend on the size of the device screen. For example, our chosen parameters may not be sufficient to protect against a keystroke inference attack on devices with a large screen since the key area is larger and the resulting modified readings may still point to the same characters.

## VIII. Related Work

Several papers have studied the practicality of motion-based keystroke inference attacks. We have discussed TapLogger [17] in Section III already. TapLogger uses the accelerometer for tap detection and the accelerometer and the orientation sensor for tap inference. The authors use k-means clustering for inferring numerical PINs and SVM for inferring passwords.

TouchLogger [11] utilizes data collected only from the orientation sensor to correctly infer more than 70% of the keys typed on a number-only on-screen keyboard. The authors use a probability density function for classification.

Accessory [16] uses the accelerometer for both tap detection and tap inference. 46 features and various machine-learning algorithms are used for classification. The presented attack can be applied to both a 6*10 keyboard and a QWERTY keyboard. It can achieve 24.5% accuracy for the 60 keyboard mode and guess 6 of 99 6-alphabets passwords in 4.5 trials on average.

Aviv et al. [10] also exclusively use the accelerometer. Their attacks pay much attention to the machine learning part and they use logistic regression and hidden Markov models. They test their attacks on a number-only keyboard and a pattern lock and achieve 43% and 73% accuracy, respectively, when sitting.

Cai and Chen [12] use both the accelerometer and the gyroscope. They do significant pre-processing on their data and their attack works on both alphabet keyboard and number pads. They can achieve 35% and 55% accuracy, respectively.

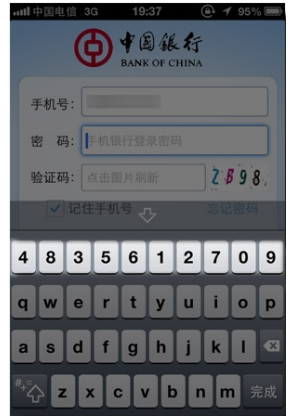

Fig. 8: An example randomized banking keyboard [4].

TapPrints [15] also uses both the accelerometer and the gyroscope to infer a key. Their attack relies on 273 features and an ensemble of machine-learning classifiers. They achieve 90% accuracy for icons and 80% for alphabets.

Al-Haiqi et al. [9] compare the results when using different motion sensors (rotation vector, accelerometer, gyroscope, and orientation sensor) and argue that the gyroscope can achieve the best accuracy.

A randomized keyboard layout cannot only be used to defend against motion-based keystroke inference attacks. Some banking apps include a simple randomized keyboard to defend against keylogging attacks. Figure 8 gives an example.

## IX. Future Work

An alternative approach to reducing the accuracy of sensor data may be reducing its sampling rate (see Section IV-C. We have simulated this approach using our gathered data and find that when halving the sampling rate, the top 4 inference accuracy will decrease to 33%, which is just as good as random guess. However, as mentioned in Section IV-C, other researchers have significant concerns about the negative impact of this approach on other apps so we need to run additional experiments to investigate these concerns.

As mentioned in Section IV-B, displaying a random keyboard layout is basically a substitution cipher, which is vulnerable to frequency analysis attack. We need to find a good frequency to shuffle the layout. It is a trade off between usability and security. We also need to investigate the possibility of limiting the amount of randomization. For example, we could shuffle a key only with keys in its proximity to keep the keyboard more familiar to a user.

Our evaluation has shown that the keyboard layout should not be randomized when long text is entered or it should be randomized only temporarily while a piece of sensitive information being part of this text is entered. The latter approach requires a way for the system to detect when sensitive information is input or alternately a way for the user to signal to the system when the user is entering sensitive information.

As mentioned in Section IV-A, existing work uses the accelerometer for tap detection. We leave it to future work to study whether other sensors that require no permissions, such as the gyroscope, could also be used for tap detection.

## X. Conclusion

With the increasing power of smartphones, attacks on smartphones also become more powerful. Motion-based keystroke inference attacks are an example. We propose two defenses against such attacks. First, we modify the Android operating system to reduce the accuracy of sensor readings. Second, we develop a randomized keyboard layout wherein the numbers and alphabets are randomized whenever the keyboard pops up. We also conduct a user study on 10 users. We find that our modifications to Android have no impact on a game that requires the accelerometer and on a pedometer. Users even prefer our modified system to the unmodified one. The randomized keyboard layout significantly increases the time it takes to enter some input. However, the participants indicate that they are willing to use such a layout while inputting sensitive information.

## Acknowledgements

We thank the anonymous reviewers for their helpful comments. This work is supported by a Google Focused Research Award, the Ontario Research Fund, and the Natural Sciences and Engineering Research Council of Canada.